\documentclass[prd,superscriptaddress,unsortedaddress,twocolumn,showpacs,preprintnumbers,amsmath,amssymb]{revtex4}

\usepackage[dvips,dvipdfmx]{graphicx}
\usepackage{amsmath,amssymb,times}
\usepackage{braket}

\def\fun#1#2{\lower3.6pt\vbox{\baselineskip0pt\lineskip.9pt
\ialign{$\mathsurround=0pt#1\hfil##\hfil$\crcr#2\crcr\sim\crcr}}}

\newcommand{\beq}{\begin{equation}}
\newcommand{\eeq}{\end{equation}}
\newcommand{\bea}{\begin{eqnarray}}
\newcommand{\eea}{\end{eqnarray}}

\DeclareSymbolFont{boldletters}{OML}{cmm} {b}{it}
\DeclareSymbolFontAlphabet{\mathbit}{boldletters}
\DeclareMathSymbol{\alpha}{\mathalpha}{letters}{"0B}
\DeclareMathSymbol{\beta}{\mathalpha}{letters}{"0C}
\DeclareMathSymbol{\gamma}{\mathalpha}{letters}{"0D}
\DeclareMathSymbol{\delta}{\mathalpha}{letters}{"0E}
\DeclareMathSymbol{\epsilon}{\mathalpha}{letters}{"0F}
\DeclareMathSymbol{\zeta}{\mathalpha}{letters}{"10}
\DeclareMathSymbol{\eta}{\mathalpha}{letters}{"11}
\DeclareMathSymbol{\theta}{\mathalpha}{letters}{"12}
\DeclareMathSymbol{\iota}{\mathalpha}{letters}{"13}
\DeclareMathSymbol{\kappa}{\mathalpha}{letters}{"14}
\DeclareMathSymbol{\lambda}{\mathalpha}{letters}{"15}
\DeclareMathSymbol{\mu}{\mathalpha}{letters}{"16}
\DeclareMathSymbol{\nu}{\mathalpha}{letters}{"17}
\DeclareMathSymbol{\xi}{\mathalpha}{letters}{"18}
\DeclareMathSymbol{\pi}{\mathalpha}{letters}{"19}
\DeclareMathSymbol{\rho}{\mathalpha}{letters}{"1A}
\DeclareMathSymbol{\sigma}{\mathalpha}{letters}{"1B}
\DeclareMathSymbol{\tau}{\mathalpha}{letters}{"1C}
\DeclareMathSymbol{\upsilon}{\mathalpha}{letters}{"1D}
\DeclareMathSymbol{\phi}{\mathalpha}{letters}{"1E}
\DeclareMathSymbol{\chi}{\mathalpha}{letters}{"1F}
\DeclareMathSymbol{\psi}{\mathalpha}{letters}{"20}
\DeclareMathSymbol{\omega}{\mathalpha}{letters}{"21}
\DeclareMathSymbol{\varepsilon}{\mathalpha}{letters}{"22}
\DeclareMathSymbol{\vartheta}{\mathalpha}{letters}{"23}
\DeclareMathSymbol{\varpi}{\mathalpha}{letters}{"24}
\DeclareMathSymbol{\varrho}{\mathalpha}{letters}{"25}
\DeclareMathSymbol{\varsigma}{\mathalpha}{letters}{"26}
\DeclareMathSymbol{\varphi}{\mathalpha}{letters}{"27}
\DeclareMathSymbol{\Gamma}{\mathalpha}{letters}{"00}
\DeclareMathSymbol{\Delta}{\mathalpha}{letters}{"01}
\DeclareMathSymbol{\Theta}{\mathalpha}{letters}{"02}
\DeclareMathSymbol{\Lambda}{\mathalpha}{letters}{"03}
\DeclareMathSymbol{\Xi}{\mathalpha}{letters}{"04}
\DeclareMathSymbol{\Pi}{\mathalpha}{letters}{"05}
\DeclareMathSymbol{\Sigma}{\mathalpha}{letters}{"06}
\DeclareMathSymbol{\Upsilon}{\mathalpha}{letters}{"07}
\DeclareMathSymbol{\Phi}{\mathalpha}{letters}{"08}
\DeclareMathSymbol{\Psi}{\mathalpha}{letters}{"09}
\DeclareMathSymbol{\Omega}{\mathalpha}{letters}{"0A}

\begin{document}

%%%%%%%%%%%%%%%%%%%%
%%%%% AUTHORS %%%%%
%%%%%%%%%%%%%%%%%%%%
\title{Determination of hadron-quark phase transition line
\\
from lattice QCD and two-solar-mass neutron star observations}
\author{Junpei Sugano}
\email[]{sugano@phys.kyushu-u.ac.jp}
\affiliation{Department of Physics, Graduate School of Sciences, Kyushu University,
             Fukuoka 812-8581, Japan}             

\author{Hiroaki Kouno}
\email[]{kounoh@cc.saga-u.ac.jp}
\affiliation{Department of Physics, Saga University,
             Saga 840-8502, Japan}  

\author{Masanobu Yahiro}
\email[]{yahiro@phys.kyushu-u.ac.jp}
\affiliation{Department of Physics, Graduate School of Sciences, Kyushu University,
             Fukuoka 812-8581, Japan}

\date{\today}

%%%%%%%%%%%%%%%%%%%%%%%%%%%%%%%%%%%%%%%%%%%%%%%%%%%%%%%%%%%%%%%%%%%%%%%%%%%
%%%%% ABSTRACT
%%%%%%%%%%%%%%%%%%%%%%%%%%%%%%%%%%%%%%%%%%%%%%%%%%%%%%%%%%%%%%%%%%%%%%%%%%%
\begin{abstract}
We aim at drawing the hadron-quark phase transition line
in the QCD phase diagram by using
the two phase model (TPM) in which
the entanglement Polyakov-loop extended Nambu--Jona-Lasinio (EPNJL) model
with vector-type four-quark interaction is used for the quark phase and
the relativistic mean field (RMF) model is for the hadron phase.
Reasonable TPM is constructed
by using lattice QCD data and neutron star observations
as reliable constraints.
For the EPNJL model, we determine the strength of
vector-type four-quark interaction at zero quark chemical potential
from lattice QCD data on quark number density normalized by its
Stefan-Boltzmann limit.
For the hadron phase, we consider
three RMF models, NL3, TM1 and model proposed by Maruyama,
Tatsumi, Endo and Chiba (MTEC).
We find that MTEC is most consistent with the
neutron star observations and TM1 is the second best.
Assuming that the hadron-quark phase transition occurs
in the core of neutron star,
we explore the density-dependence of
vector-type four-quark interaction.
Particularly for the critical baryon chemical potential
$\mu^{\rm c}_{\rm B}$ at zero temperature,
we determine a range of $\mu^{\rm c}_{\rm B}$
for the quark phase to occur in the core of neutron star. 
The values of $\mu^{\rm c}_{\rm B}$ lays in the range
$1560 \textrm{MeV} \le \mu^{\rm c}_{\rm B}\le 1910$ MeV.
\end{abstract}

\maketitle

%%%%%%%%%%%%%%%%%%%%%%%%%%%%%%%%%%%%%%%%%%%%%%%%%%%%%%%%%%%%%%%%%%%%%%%%%%%
%%%%%  Introduction 
%%%%%%%%%%%%%%%%%%%%%%%%%%%%%%%%%%%%%%%%%%%%%%%%%%%%%%%%%%%%%%%%%%%%%%%%%%%
\section{INTRODUCTION}
Temperature ($T$) and baryon chemical potential ($\mu_{\rm B}$)
dependence of quantum chromodynamics (QCD)
is often described as the QCD phase diagram~\cite{bib1},
where $\mu_{\rm B}$ is related to quark chemical
potential $\mu_{\rm q}$ as $\mu_{\rm B}=3\mu_{\rm q}$.
Investigation of the truth about the QCD phase diagram
is quite important not only in
hadron physics but also in particle physics and astrophysics.
Lattice QCD (LQCD) simulation as the first principle calculation is
a powerful tool of studying the QCD phase diagram.
In fact, recent LQCD simulations provide reliable results in
$\mu_{\rm q}/T\lesssim 1$ with sophisticated methods
~\cite{bib2,bib3,bib4,bib5,bib6,bib7,bib8,bib9,bib10,bib11}.
However, these methods are considered not to work well
in $\mu_{\rm q}/T\gtrsim 1$ because of the severe sign problem.
To understand the QCD phase diagram there,
many effective models were proposed so far.
Among the effective models,
the entanglement Polyakov-loop extended Nambu--Jona-Lasinio (EPNJL) model
is one of the most useful effective models~\cite{bib17}.
The 2-flavor EPNJL model is successful in reproducing
LQCD data at zero and imaginary $\mu_{\rm q}/T$,
isospin chemical potential and small real $\mu_{\rm q}/T$~\cite{bib17,bib18}.
In addition, Ishii \textit{et.al.} showed very recently that
random-phase-approximation calculations based on the EPNJL model
well reproduce $T$ dependence of the meson screening masses
calculated by LQCD in both the 2- and 2+1-flavor cases~\cite{bib19,bib20}.

In spite of the success, the EPNJL model can not treat the 
baryon degrees of freedom explicitly. This is a disadvantage
of the EPNJL model in describing the baryon sector in the QCD phase diagram.
Another way of describing all the region of QCD phase diagram
is the two phase model (TPM) in which
the hadron-quark phase transition is assumed to be the first order
and the phase boundary is determined by the Gibbs criterion~\cite{bib21,bib22}.
The TPM allows us to use different models for hadron and quark phases.
Various methods were proposed and developed so far 
to describe the hadron phase; for example,
the Brueckner-Hartree-Fock method~\cite{bib23},
its relativistic version~\cite{bib24}, the variational method~\cite{bib25}
and the relativistic mean field (RMF) model~\cite{bib26}.
Among them, we use the RMF model in this paper since it is easy to treat
and successful in describing the saturation properties of the nuclear matter.
However, the equation of state (EoS) strongly
depend on the choice of parameters and are quite different,
especially above the normal nuclear density $\rho_{0}$.
Observations of neutron star (NS) may be a key to solve this problem.
Recently, two-solar-mass ($2M_{\rm sun}$) NSs
were discovered with high accuracy~\cite{bib27,bib28},
and Steiner \textit{et.al.} yielded the best
fitting against various observed mass-radius (MR) relations~\cite{bib29}.
Therefore, we can judge what version of the RMF model
is most reasonable above $\rho_{0}$
because MR relation is sensitive to the EoS taken.

In the core of heavy NSs,
it is possible that the hadron-quark phase transition takes place.
The occurrence of the transition depends on
stiffness of quark-phase EoS,
which is sensitive to the strength $G_{\rm v4}$ of the vector-type
four-quark interaction in the EPNJL model.
In our previous work~\cite{bib30}, the value of $G_{\rm v4}$ at
$\mu_{\rm q}/T=0$ was determined
from LQCD data on the quark number density $n_{\rm q}$
normalized by its Stefan-Boltzmann limit $n_{\rm SB}$;
note that $n_{\rm q}/n_{\rm SB}$ is $\mu_{\rm q}$-even
and has no finite-volume effect.
The value of $G_{\rm v4}$ obtained in the $\mu_{\rm q}/T=0$ limit
is called $G_{\rm v4}(0)$ in the present paper.
As for $n_{\rm q}/n_{\rm SB}$, new LQCD data on $n_{\rm q}$
was provided by using the extrapolation
from the imaginary $\mu_{\rm q}/T$ region
to the real one~\cite{bib11}.
Since LQCD simulations in the imaginary $\mu_{\rm q}/T$ region are free
from the sign problem,
the numerical errors of the new data are very small
compared with the previous one based on the Taylor expansion method at
real $\mu_{\rm q}/T$~\cite{bib4}.
This suggests that one can determine
the value of $G_{\rm v4}(0)$ more sharply.

If the strength $G_{\rm v4}$ is decreasing with increasing the $\mu_{\rm q}/T$,
the possibility that the quark phase exist in the core
of NS becomes higher. 
However, at present, it is difficult to determine the density-dependence of $G_{\rm v}$ theoretically. 
Hence, here, we consider an inverse problem.  
When we assume the existence of the quark phase in the core of NS, 
how does the existence constrain the  density-dependence of 
the strength $G_{\rm v4}$?
How much should the critical baryon chemical potential of
hadron-quark phase transition be?

In this paper, we first construct reasonable TPMs by
using LQCD data at $\mu_{\rm q}/T=0$ as a constraint
on quark-phase EoS and NS observations
as a constraint on both hadron- and quark-phase EoS.
As a quark part of TPM, we consider three types of EPNJL models;
(1) the model with no vector-type four-quark interaction,
(2) the model with vector-type four-quark interaction
in which the strength $G_{\rm v4}$ is assumed to be constant,
i.e., $G_{\rm v4}=G_{\rm v4}(0)$,
and
(3) the model with the vector-type four-quark interaction
in which the density-dependent strength $G_{\rm v4}(n_{\rm q})$
is introduced.
The value of $G_{\rm v4}(0)$ is determined from LQCD data
on $n_{\rm q}/n_{\rm SB}$ in the $\mu_{\rm q}/T=0$ limit.
The density dependence of $G_{\rm v4}(n_{\rm q})$ is
discussed by assuming that the quark phase takes place in
the core of NS.
As hadron phase models, we take three RMF models, i.e., 
TM1~\cite{bib31}, NL3~\cite{bib32} and the model proposed 
by Maruyama, Tatsumi, Endo and Chiba (MTEC)~\cite{bib33}. 
We determine which hadron-phase EoS
is consistent with $2M_{\rm sun}$ NS observations
and the statistically analyzed MR relation by Steiner
\textit{et.al.}~\cite{bib27,bib28,bib29}.
We focus our attention on the $2M_{\rm sun}$ region
on the statistically analyzed MR relation,
since our interest is whether the hadron-quark
phase transition takes place or not in the core of NS
and this possibility becomes higher for heavy NS.
We will find that MTEC EoS well reproduces
all the data on MR relation, particularly in the $2M_{\rm sun}$ region.
The second best is TM1 EoS.

We then pick up MTEC and TM1 as hadron-phase EoSs and
consider six types of TPMs, as shown in TABLE~\ref{tb:table1}.
These are classified with the hadron-phase EoS,
that is, MTEC EoS as a TPMa and TM1 EoS as a TPMb.
For each class, we take EPNJL of type (1)--(3)
for the quark-phase EoS.
By using TPMa1--TPMa3 and TPMb1--TPMb3,
we calculate the MR relation and
draw the hadron-quark phase transition line
in the $T$-$\mu_{\rm B}$ plane.
For TPMa3 and TPMb3,
varying $n_{\rm q}$ dependence of $G_{\rm v4}(n_{\rm q})$,
we determine the upper bound of the transition line
for the quark phase to appear in the core of NS.

The paper is organized as follow:
In Sec. \ref{sec2},
we formulate the EPNJL model
with vector-type four-quark interaction and the RMF model.
The prescription of the Gibbs criterion is also explained.
Sec. \ref{sec3} is devoted to show the numerical results.
We first determine the value of $G_{\rm v4}(0)$ by using
new LQCD data on $n_{\rm q}/n_{\rm SB}$ in the $\mu_{\rm q}/T=0$ limit.
Next, we select the RMF model through the comparison with
the data on MR relation.
Finally, we construct the TPMa1--TPMa3 and TPMb1--TPMb3.
From these models,
we draw the upper and lower bounds of
hadron-quark phase transition line from
the condition that the quark phase takes place in the core of NS.
The density-dependence of the vector-type four-quark interaction
is also discussed.

%%%%%%%%%%%%%%%%%%%%%%%%%%%%%%%%%%%%%%%%%%%%%%%%%%%%%%%%%
%%%%% TABLE 1 %%%%%%%%%%%%%%%%%%%%%%%%%%%%%%%%%%%%%%%%%%%
%%%%%%%%%%%%%%%%%%%%%%%%%%%%%%%%%%%%%%%%%%%%%%%%%%%%%%%%%
\begin{table}[t]
\begin{center}
\caption{TPMs taken in this paper.
The TPMs are combinations 
of RMF model (MTEC or TM1) and three EPNJL models 
of types (1)--(3).
See the text for the definition of RMF and EPNJL models. 
   }
\begin{tabular}{c|c|c|c}
\hline \hline
class & hadron-phase EoS & quark-phase EoS & label \\ \hline
     &  & EPNJL of type (1) & TPMa1 \\
TPMa & MTEC & EPNJL of type (2) & TPMa2 \\
     &  & EPNJL of type (3) & TPMa3 \\ \hline
     &  & EPNJL of type (1) & TPMb1 \\
TPMb & TM1 & EPNJL of type (2) & TPMb2 \\
     &  & EPNJL of type (3) & TPMb3 \\ \hline \hline
\end{tabular}
\label{tb:table1}
\end{center}
\end{table}
%%%%%%%%%%%%%%%%%%%

%%%%%%%%%%%%%%%%%%%%%%%%%%%%%%%%%%%%%%%%%%%%%%%%%%%%%%%%%%%%%%%%%%%%%%%%%%%
%%%%% FORMULATION
%%%%%%%%%%%%%%%%%%%%%%%%%%%%%%%%%%%%%%%%%%%%%%%%%%%%%%%%%%%%%%%%%%%%%%%%%%%
\section{MODEL SETTING}
\label{sec2}

%%%%%%%%%%%%%%%%%%%%%%%%
\subsection{QUARK PHASE}
%%%%%%%%%%%%%%%%%%%%%%%%
The Lagrangian of the EPNJL of type (1) is given by
%%%%%
\begin{align}
{\cal L}_{\textrm{EPNJL}}
= &{\bar q}(i\gamma^\mu D_\mu - m_0)q - {\cal U}(\Phi,{\Phi}^{\ast})
\notag \\
  & + \tilde{G}_{\rm s4}[({\bar q}q )^2  + ({\bar q }i\gamma_5 {\vec \tau}q)^2]
     - \tilde{G}_{\rm v4}(0)({\bar q \gamma_\mu q})^2
\label{eq:EPNJL_gv0}
\end{align}
%%%%%
where $q=(u,d)^{\textrm{T}}$ is u- and d- quark fields,
$m_{0}=\textrm{diag}(m_{\textrm{u}},m_{\textrm{d}})$
denotes a current quark mass matrix and $\vec{\tau}$ is an isospin-matrix.
In this paper, we set $m_{\textrm{u}}=m_{\textrm{d}}\equiv m_{0}$.
The quark and gluon interact through the covariant derivative
$D^\mu=\partial^\mu+iA^\mu$,
where $A^\mu=g\delta^{\mu}_{0}A^0_{a}{\lambda_a/2}
=-ig\delta^{\mu}_{0}(A_{4})_a{\lambda_a/2}$ with gauge field $A^\mu_a$,
Gell-Mann matrix $\lambda_a$ and the gauge coupling $g$.
$\tilde{G}_{\rm s4}$ and $\tilde{G}_{\rm v4}(0)$ are
the strength of scalar- and vector-type four-quark interactions
depending on the Polyakov loop $\Phi$ and its conjugate $\Phi^{\ast}$.
We parametrize the Polyakov-loop dependence of these interactions as 
%%%%%
\begin{align*}
 &\tilde{G}_{\rm s4} = G_{\rm s4}\left(1-\alpha_1\Phi{\Phi}^* -\alpha_2(\Phi^3 + {\Phi^*}^{3})\right)
\\
 &\tilde{G}_{\rm v4}(0) = G_{\rm v4}(0)\left(1-\alpha_1\Phi{\Phi}^* -\alpha_2(\Phi^3 + {\Phi^*}^{3})\right)
\end{align*}
%%%%%
according to the previous works~\cite{bib17,bib30}.

Eventually, the NJL sector of Eq. (\ref{eq:EPNJL_gv0}) has five parameters 
$(m_{0}, G_{\rm s4},G_{\rm v4}(0), \alpha_1, \alpha_2)$.
We take $G_{\rm s4}=5.498\ \rm{GeV}^{-2}$ and $\alpha_{1}=\alpha_{2}=0.2$
of Ref. \cite{bib17}.
The value of $G_{\rm v4}(0)$ will be determined from LQCD data
on $n_{\rm q}/n_{\rm SB}$~\cite{bib4,bib11}.
In the LQCD data we use, the corresponding
current quark mass $m_{0}$ was 130 MeV and
it is much heavier than the empirical value $\sim$ 5 MeV. 
LQCD simulations were done
by the Taylor expansion method~\cite{bib4}
and the imaginary $\mu_{\rm q}/T$ method~\cite{bib11}. 
The two kinds of simulations used
2-flavor clover-improved Wilson fermion along the line of 
constant physics of $m_{\pi}/m_{\rho}=0.8$ for 
$\pi$- and $\rho$-meson masses $m_{\pi}$ and $m_{\rho}$.
We also keep $m_{0}=130$ MeV for our EPNJL model analysis
to determine the value of $G_{\rm v4}(0)$ from the LQCD data.

In the EPNJL model, only the time component of 
gluon field $A^{\mu}_{a}$ is treated as
a homogeneous and static background field.
We define $\Phi$ and $\Phi^{\ast}$ in the Polyakov gauge as 
%%%%%
\begin{align}
 \Phi = \frac{1}{3}{\rm Tr}_{\rm c}(L),\ \ \
 \Phi^{\ast} = \frac{1}{3}{\rm Tr}_{\rm c}({L^\dag}) ,
\end{align}
%%%%%
where $L= \exp[i A_4/T]=\exp[i~{\rm diag}(A_4^{11},A_4^{22},A_4^{33})/T]$ 
for the classical variables $A_4^{ii}$ 
satisfying $A_4^{11}+A_4^{22}+A_4^{33}=0$. 
Under the definition eq. (2),
we use the logarithm-type Polyakov potential $\mathcal{U}(\Phi,\Phi^{\ast})$
proposed in Ref.~\cite{bib34},
%%%%%
\begin{align}
 &\mathcal{U}(\Phi,\Phi^{\ast})=T^4\left[-\frac{a(T)}{2}\Phi\Phi^{\ast}+b(T)\log H(\Phi,\Phi^{\ast})\right],
\end{align}
%%%%%
where
%%%%%
\begin{align*}
 &a(T)=a_{0}+\left(\frac{T_{0}}{T}\right)+a_{2}\left(\frac{T_{0}}{T}\right)^2,\
 \ b(T)=b_{3}\left(\frac{T_{0}}{T}\right)^3 \\
 &H(\Phi,\Phi^{\ast})=1-6\Phi\Phi^{\ast}+4(\Phi^3+\Phi^{\ast 3})-3(\Phi\Phi^{\ast})^2.
\end{align*}
%%%%%
Usually, the parameter $T_{0}$ is 270 MeV so as to reproduce
LQCD data in the pure gauge limit~\cite{bib35}.
For this value of $T_{0}$, however,
the EPNJL model yields a larger value of
pseudo-critical temperature $T_\textrm{pc}$ for the deconfinement
transition than the full-LQCD prediction
171 MeV at $\mu_{\rm q}/T=0$~\cite{bib36,bib37,bib38}.
We then rescale $T_0$ to 190 MeV. By this rescale,
the EPNJL model reproduce $T_{\rm pc} = 171$ MeV.
Other parameters $(a_{0}, a_{1}, a_{2}, b_{3})$ are summarized in TABLE~\ref{tb:table2}.

%%%%%%%%%%%%%%%%%%%%%%%%%%%%%%%%%%%%%%%%%%%%
%%%%% TABLE 2 %%%%%%%%%%%%%%%%%%%%%%%%%%%%%%
%%%%%%%%%%%%%%%%%%%%%%%%%%%%%%%%%%%%%%%%%%%%
\begin{table}[t]
\begin{center}
\caption{The parameter set in the Polyakov potential proposed in
 Ref.~\cite{bib34}. All parameters are dimensionless.}
\begin{tabular}{c|c|c|c}
\hline \hline
$a_{0}$ & $a_{1}$ & $a_{2}$ & $b_{3}$ \\ \hline
\ \ \ \ \ 3.51\ \ \ \ \  & \ \ \ \ \ $-2.47$\ \ \ \ \  & 
\ \ \ \ \ 15.2\ \ \ \ \  & \ \ \ \ \ $-1.75$\ \ \ \ \  \\ \hline \hline
\end{tabular}
\label{tb:table2}
\end{center}
\end{table}

After the mean field approximation to Eq. (\ref{eq:EPNJL_gv0}),
one can obtain the thermodynamic potential $\Omega_{\rm EPNJL}$ (per unit volume) as
%%%%%
\begin{align}
&\Omega_{\rm EPNJL}
= U_{\rm M}+{\cal U}-2 \sum_{i=\textrm{u, d}} \int \frac{d^3 \mathbf{p}}{(2\pi)^3}
   \Bigl[ 3 E_{i} \notag \\
&+ \frac{1}{\beta}
           \log \left(1 + 3(\Phi+{\Phi}^* e^{-\beta (E-\tilde{\mu}_i )}) 
           e^{-\beta (E-\tilde{\mu}_{i})}+ e^{-3\beta (E-\tilde{\mu}_{i})}\right)
\notag\\
&+ \frac{1}{\beta} 
           \log \left(1 + 3({\Phi}^*+{\Phi e^{-\beta (E+\tilde{\mu}_{i})}}) 
              e^{-\beta (E+\tilde{\mu}_{i})}+ e^{-3\beta (E+\tilde{\mu}_{i})}\right)
	      \Bigl],
\end{align}
%%%%%
where $\beta=1/T$, $U_{\rm M}=\tilde{G}_{\rm s4}\sigma^2-\tilde{G}_{\rm v4}(0)n^2_{\rm q}$,
$E=\sqrt{{\bf p}^2+M^2}$ with the constituent quark mass $M=m_0-2\tilde{G}_{\rm s4}\sigma$
and $\tilde{\mu}_{\rm i}=\mu_{\rm i}-2\tilde{G}_{\rm v4}(0)n_{\rm q}$
for $i=$ u, d.
The chiral condensate and the quark numbder density are defined by
$\sigma =\langle\bar{q}q\rangle$, $n_{\rm q}=\langle q^{\dag}q\rangle$.
We use the three-dimensional momentum cutoff $\Lambda=$ 631.5 MeV
to regularize the vacuum term.
The variables $X=\sigma, n_{\textrm{q}}, \Phi, \Phi^{\ast}$ are determined with
stationary condition ${\partial \Omega_{\textrm{EPNJL}}}/{\partial X}=0$.
In this paper, we employ the approximation $\Phi=\Phi^{\ast}$ since
it is known to be good approximation~\cite{bib17}.

In the EPNJL of type (3), the density-dependent
strength $G_{\rm v4}(n_{\rm q})$ of vector-type four-quark interaction is
introduced. The strength is assumed to be a Gaussian form of
%%%%%
\begin{eqnarray}
G_{\rm v4}(n_{\rm q})=\textrm{e}^{-b\left(\frac{n_{\rm q}}{\rho_{0}}\right)^2}G_{\rm v4}(0),
\label{eq:density-dependence}
\end{eqnarray}
%%%%%
where $b$ is a parameter and $\rho_{0}$ is a saturation density.
Note that the model with vanishing  (constant) vector interaction coupling is obtained when 
$b \to \infty$ ($b\to 0$).
The thermodynamic potential of EPNJL of type (3) can be obtained by
the replacement $G_{\rm v4}(0)\rightarrow G_{\rm v4}(n_{\rm q})$.
The detail will be discussed in Sec. \ref{sec3}.

%%%%%%%%%%%%%%%%%%%%%%%%%%%%%%%%%%%%%%%%%%%
\subsection{RELATIVISTIC MEAN FIELD MODEL}
%%%%%%%%%%%%%%%%%%%%%%%%%%%%%%%%%%%%%%%%%%%

%%%%%%%%%%%%%%%%%%%%%%%%%%%%%%%%%%%%%%%%%%%%%%%%
%%%%% TABLE 3 %%%%%%%%%%%%%%%%%%%%%%%%%%%%%%%%%%
%%%%%%%%%%%%%%%%%%%%%%%%%%%%%%%%%%%%%%%%%%%%%%%%
\begin{table}[t]
\begin{center}
\caption{Three parameter sets used in the RMF models.
The saturation properties derived from three parameter sets are
also summarized.
Shown are the saturation density $\rho_{0}$, binding energy $E_{0}$,
incompressibility $K$, symmetry energy $S_{0}$, and ratio of the
effective nucleon mass $M_{\rm N}$ to nucleon mass $m_{\rm N}$.}
\begin{tabular}{cccc}
\hline \hline
parameter & MTEC & TM1 & NL3 \\ \hline
$m_{\textrm{N}}$ (MeV) & 938 & 938 & 939 \\
$m_{\varphi}$ (MeV) & 400 & 511.198 & 508.194 \\ 
$m_{\omega}$ (MeV) & 783 & 783 & 782.501 \\ 
$m_{\rho}$ (MeV) & 769 & 770 & 763 \\
$g_{\varphi}$ & 6.3935 & 10.0289 & 10.217 \\
$g_{\omega}$ & 8.7207 & 12.6139 & 12.868 \\
$g_{\rho}$ & 4.2696 & 4.6322 & 4.474 \\
$g_{2}$ ($\textrm{fm}^{-1}$) & $- 10.757$ & $- 7.2325$ & $- 10.431$ \\
$g_{3}$ & $- 4.0452$ & 0.6183 & $- 28.885$ \\
$c_{3}$ & 0 & 71.3075 & 0 \\ \hline \hline
saturation property & MTEC & TM1 & NL3 \\ \hline
$\rho_{0}$ ($\textrm{fm}^{-3}$) & 0.153 & 0.145 & 0.148 \\
$E_{0}$ (MeV) & $- 16.3$ & $- 16.3$ & $- 16.3$ \\ 
$K$ (MeV) & 240 & 281 & 271 \\ 
$S_{0}$ (MeV) & 32.5 & 36.9 & 37.4 \\
$M_{\rm N}/m_{\rm N}$ & 0.78 & 0.63 & 0.60 \\ \hline \hline
\end{tabular}
\label{tb:table3}
\end{center}
\end{table}

% ----- Fig1 -----
\begin{figure*}[]%[H]
\begin{center}
\includegraphics[width=0.8\textwidth]{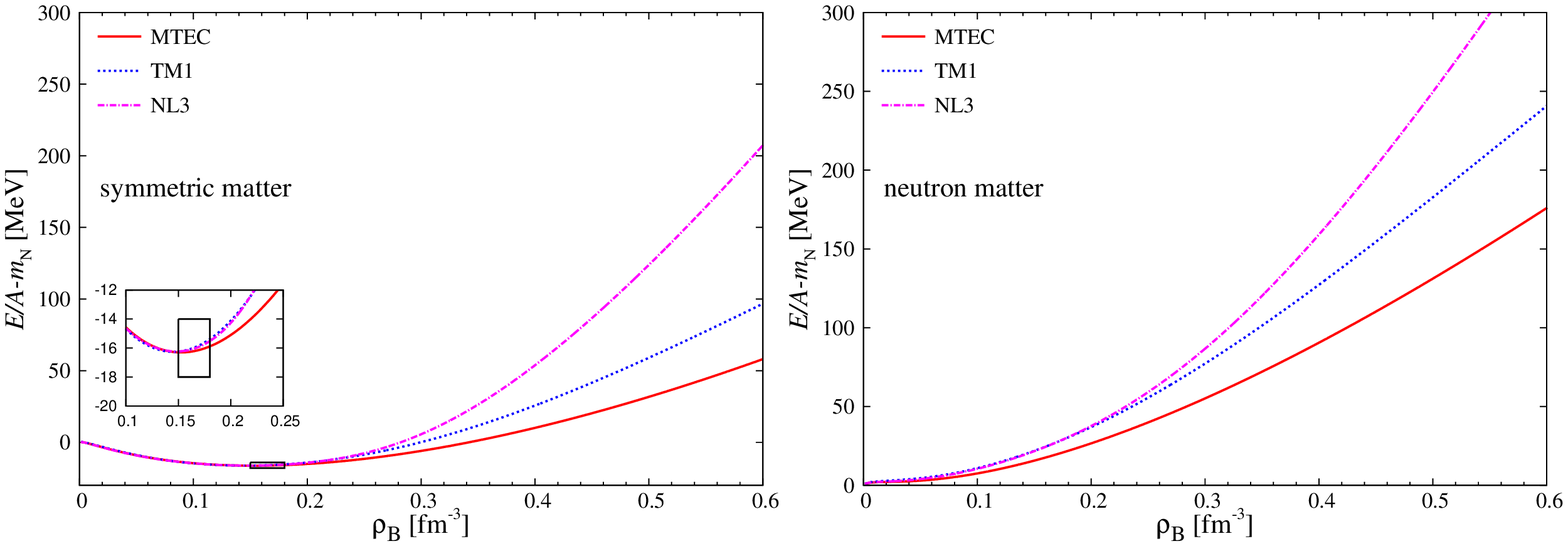}
\end{center}
\caption{
The equation of states of three parameter sets
for symmetric matter (left panel) and neutron matter (right panel).
In both panels, the results of MTEC, TM1 and NL3
are corresponding the solid, the dotted and
the dot-dashed lines, respectively.
In the left panel,
the open square indicates the empirical saturation point~\cite{bib23}.
}
\label{Fig1}
\end{figure*}
% ----------------

We treat the hadron phase by the RMF model.
In the RMF model,
the nucleon-nucleon interaction is mediated by scalar ($\varphi$),
vector ($\omega$) and isovector ($\rho$) mesons.
The Lagrangian of RMF model is written as
%%%%%
\begin{align}
 \mathcal{L}_{\textrm{RMF}}=&\bar{\psi}(i\gamma^{\mu}\partial_{\mu}-m_{\textrm{N}}-g_{\varphi}\varphi
                   -g_{\omega}\gamma^{\mu}\omega_{\mu}-g_{\rho}\gamma^{\mu}\rho^{a}_{\mu}\tau_{a})\psi
 \notag \\
                  &+\frac{1}{2}\partial^{\mu}\varphi\partial_{\mu}\varphi-\frac{1}{2}m^2_{\varphi}\varphi^2-\frac{1}{3}g_{2}\varphi^3
                   -\frac{1}{4}g_{3}\varphi^4
 \notag \\
                  &-\frac{1}{4}\Omega^{\mu\nu}\Omega_{\mu\nu}+\frac{1}{2}m^2_{\omega}\omega^{\mu}\omega_{\mu}
                   +\frac{1}{4}c_{3}(\omega^{\mu}\omega_{\mu})^2
 \notag \\
                  &-\frac{1}{4}R^{\mu\nu}_{a}R_{\mu\nu}^{a}+\frac{1}{2}m^2_{\rho}\rho^{\mu}_{a}\rho^{a}_{\mu},
\end{align}
%%%%%
where $\psi$ is the nucleon (N) field, and
$\Omega^{\mu\nu}$ ($R^{\mu\nu}_{a}$)
is the field strength of $\omega$ ($\rho$) meson.
Masses of the particles are denoted by $m_{\rm N}, m_{\varphi}, m_{\omega}, m_{\rho}$,
Yukawa-coupling constants of nucleon with mesons
are by $g_{\varphi}, g_{\omega}, g_{\rho}$ and
self-interactions of $\varphi$ and
$\omega$ mesons are by $g_{2}, g_{3}$ and $c_{3}$.
We take three RMF models of TM1~\cite{bib31}, NL3~\cite{bib32} and MTEC~\cite{bib33}.
The parameter sets of three models
are summarized in TABLE~\ref{tb:table3},
together with the saturation properties
calculated by the models.

Under the mean field approximation,
all the meson fields $\varphi$, $\omega$, $\rho$ are replaced by the mean values
$\langle \varphi \rangle$,
$\langle \omega^{0}\rangle\delta^{\mu 0}$,
$\langle \rho^{0}_{3}\rangle\delta^{\mu 0}\delta_{a 3}$, respectively.
For simplicity, these 
mean values are denoted by $\varphi$, $\omega$, $\rho$.
The mean values are determined by the Euler-Lagrange equations,
%%%%%
\begin{align}
 & m^2_{\varphi}\varphi+g_{2}\varphi^2+g_{3}\varphi^3=-g_{\varphi}\rho_{\textrm{s}},
\\
 & m^2_{\omega}\omega+c_{3}\omega^3=g_{\omega}\rho_{\rm B}, 
\\
 & \rho=\frac{g_{\rho}}{m^2_{\rho}}\rho_{\rm I},
\end{align}
%%%%%
where $\rho_{\textrm{s}}$, $\rho_{\textrm{B}}$, $\rho_{\textrm{I}}$
are scalar, baryon-number and isospin densities.

The thermodynamic potential of the RMF model $\Omega_{\rm RMF}$ (per unit volume) is
then obtained by
%%%%%
\begin{align}
 &\Omega_{\rm RMF}=U_{\rm meson}-\frac{2}{\beta}\sum_{i=\textrm{p,n}}
 \int \frac{d^3 \mathbf{p}}{(2\pi)^3}
\notag \\
 &\times \Bigl[ \log(1+\textrm{e}^{-\beta(E-\tilde{\mu}_{i})})
       +\log(1+\textrm{e}^{-\beta(E+\tilde{\mu}_{i})})
 \Bigr],
\end{align}
%%%%%
where $E=\sqrt{\mathbf{p}^2+M^2_{\textrm{N}}}$ for the
nucleon effective mass $M_{\textrm{N}}= m_{\textrm{N}}+g_{\varphi}\varphi$, and
%%%%%
\begin{align*}
 U_{\rm meson}=&
 \frac{1}{2}m^2_{\varphi}\varphi^2+\frac{1}{3}g_{2}\varphi^3+\frac{1}{4}g_{3}\varphi^4 \\
 &-\frac{1}{2}m^2_{\omega}\omega^2-\frac{1}{4}c_{3}\omega^4-\frac{1}{2}m^2_{\rho}\rho^2
\end{align*}
%%%%%
is the mesonic potential. The effective chemical potentials
for neutron (n) and proton (p) are defined by
$\tilde{\mu}_{\textrm{n,p}}=\mu_{\textrm{n,p}}-g_{\omega}\omega\pm
g_{\rho}\rho$.

Figure. 1 shows the EoSs of symmetric matter (left panel) and neutron matter
(right panel) calculated by TM1, NL3 and MTEC at $T=0$.
As for densities smaller than the saturation point (open square),
all the EoSs yield an universal line.
On the other hand,
there are remarkable differences among the EoSs
for densities higher than the saturation point.
MTEC EoS is softest, whereas NL3 EoS is stiffest.
TM1 EoS lies halfway between them.
The behavior of EoS in $\rho_{B}\gtrsim \rho_{0}$
largely affects the MR relation of NSs.
Therefore, we can select which EoS model
is preferable for the MR relation,
particularly in $2M_{\rm sun}$ region.

%%%%%%%%%%%%%%%%%%%%%%%%%%%%%
\subsection{TWO PHASE MODEL}
%%%%%%%%%%%%%%%%%%%%%%%%%%%%%
In $\mu_{\rm q}/T=0$, it is established by LQCD simulations
that the hadron-quark deconfinement transition is crossover~\cite{bib39}. 
However, the pseudo-critical temperature is well estimated by the TPM~\cite{bib21}.
We thus use the TPM and the Gibbs criterion to determine the
phase boundary of the hadron-quark phase transition
for each set of $T$ and $\mu_{\textrm{B}}$.

Pressures of the EPNJL and the RMF models are obtained by
%%%%%
\begin{align}
& P_{\textrm{EPNJL}}(\mu_{\textrm{B}},T)
= - (\Omega_{\textrm{EPNJL}}(\mu_{\textrm{B}},T) - \Omega_{\textrm{EPNJL}}(0,0)) - B, \\
& P_{\textrm{RMF}}(\mu_{\textrm{B}},T) = - \Omega_{\textrm{RMF}}(\mu_{\textrm{B}},T),
\end{align}
%%%%%
where the bag constant $B$ is introduced
in $P_{\rm EPNJL}$
to describe the difference of vacuum between the hadron and quark phases.
According to the Gibbs criterion,
the quark phase (the hadron phase) takes place
for the condition $P_{\textrm{EPNJL}}>P_{\textrm{RMF}}\ (P_{\textrm{EPNJL}}<P_{\textrm{RMF}})$.
When $B$ is 100 $\textrm{MeV}^4$, our TPM can reproduce the LQCD prediction
of $T_{\textrm{pc}}=171$ MeV of the deconfinement transition at $\mu_{\textrm{q}}/T=0$.

%%%%%%%%%%%%%%%%%%%%%%%%%%%%%%%%%%%%%%%%%%%%%%%%%%%%%%%%%%%%%%%%%%%%%%%%%%%
%%%%% RESULTS
%%%%%%%%%%%%%%%%%%%%%%%%%%%%%%%%%%%%%%%%%%%%%%%%%%%%%%%%%%%%%%%%%%%%%%%%%%%
\section{RESULTS}
\label{sec3}

We show our numerical results in this section.
We first determine the value of  $G_{\rm v4}(0)$
from LQCD data on the ratio $n_{\rm q}/n_{\rm SB}$
in the $\mu_{\rm q}/T=0$ limit~\cite{bib4,bib11}.
As for the RMF model, it is shown that MTEC and TM1 are proper EoSs,
through the comparison with the NS observations~\cite{bib27,bib28,bib29}.

Next, from the combinations of the two hadron-phase EoSs and
EPNJL type (1)--(3),
we construct TPMa1 - TPMa3, TPMb1 - TPMb3.
In the TPMa3 and TPMb3,
the density-dependent strength $G_{\rm v4}(n_{\rm q})$
of vector-type four-quark interaction is introduced.
We parametrize the density dependence
with a Gaussian form having a single parameter $b$,
shown in Eq. (\ref{eq:density-dependence}).
We determine the lower bound of $b$
assuming that the hadron-quark phase transition takes place
in the core of NS.
By using six models, the MR relation and
the band of the hadron-quark phase transition line
that allows the quark phase to exist in the core of NS are calculated.

%%%%%%%%%%%%%%%%%%%%%%%%%%%%%%%%%%%%%%%%%%%%%%%%%%%%%%%%%%%%%%%%%%%%%%%%%%%%%%%%%%%%%%%%%%%%%%%
\subsection{DETERMINATION OF THE VALUE OF $G_{\rm v4}(0)$}
%%%%%%%%%%%%%%%%%%%%%%%%%%%%%%%%%%%%%%%%%%%%%%%%%%%%%%%%%%%%%%%%%%%%%%%%%%%%%%%%%%%%%%%%%%%%%%%
% ----- Fig.2 -----
\begin{figure}[t]%[H]
\begin{center}
\includegraphics[width=0.45\textwidth]{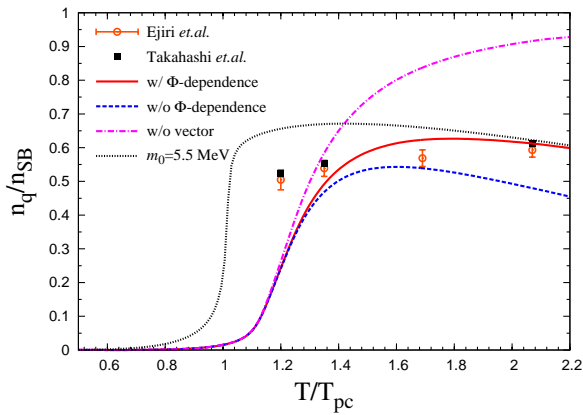}
\end{center}
\caption{
Temperature dependence of $n_{\rm q}/n_{\rm SB}$
in the $\mu_{\rm q}/T=0$ limit.
The temperature is normalized by the $T_{\rm pc}=171$ [MeV].
The data are the LQCD results~\cite{bib4, bib11}.
The lines are the results of calculations for
the case with $\tilde{G}_{\rm v4}(0)$ (solid),
$G_{\rm v4}(0)$ (dashed)
and without vector-type four-quark interaction (dot-dashed).
The dotted line corresponds to
the results with $m_{0} = 5.5$ MeV.
}
\label{Fig2}
\end{figure}
% -----------------

In the $T>T_{\rm pc}$ region,
the chiral condensate $\sigma$ is nearly equal to zero,
that is, the chiral symmetry is restored.
Hence, the scalar-type four-quark interaction becomes negligible
and only the vector-type four-quark interaction contributes to
the ratio $n_{\rm q}/n_{\rm SB}$ that is $\mu_{\rm q}$-even and therefore
finite even in the $\mu_{\rm q}/T=0$ limit.
Thus, we determine the value $G_{\rm v4}(0)$ from LQCD data
on $n_{\rm q}/n_{\rm SB}$ at $T>T_{\rm pc}$.

Figure 2 shows $T$ dependence of $n_{\rm q}/n_{\rm SB}$.
Here, $T$ is normalized by $T_{\rm pc}=171$ MeV.
In EPNJL model calculations, $m_{0}$ is taken to be 130 MeV,
as already mentioned in Sec \ref{sec2}.
If the vector-type four-quark interaction is zero,
the EPNJL model largely overestimates the LQCD data.
Meanwhile, good agreement is seen
for the case of $G_{\rm v4}(0)=0.36G_{\rm s4}$
at high $T$ such as $T=2T_{\rm pc}$.
The comparison between the solid and dashed lines suggests that
the entanglement coupling in $G_{\rm v4}(0)$
is necessary to reproduce the LQCD data.
The result of $m_{0}=5.5$ MeV is also plotted.
Comparing the dotted line with the solid line,
we find that $m_{0}$ dependence is small at high $T$.
This means that the value of $G_{\rm v4}(0)$ can be determined at high $T$
even if $m_{0}$ is heavier than physical value.

%%%%%%%%%%%%%%%%%%%%%%%%%%%%%%%%%%%%%%%%%%%
\subsection{SELECTION OF RMF MODEL}
%%%%%%%%%%%%%%%%%%%%%%%%%%%%%%%%%%%%%%%%%%%
% ----- Fig.3 -----
\begin{figure}[t]%[H]
\begin{center}
\includegraphics[width=0.45\textwidth]{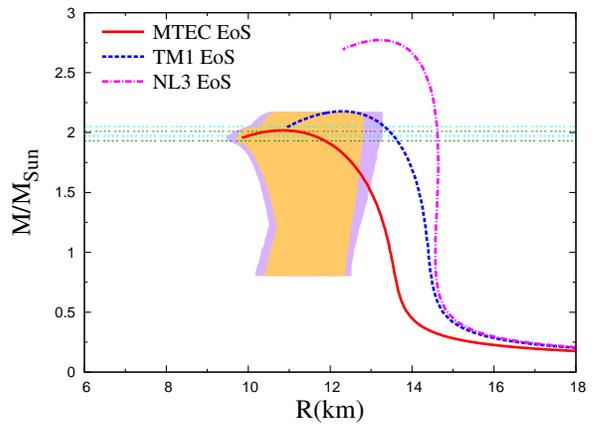}
\end{center}
\caption{
The MR relation for three RMF EoSs.
The two horizontal boxes are the
$2M_{\rm sun}$ observational data~\cite{bib27,bib28}.
The two areas correspond to the 68\%
and 95\% confidence counters
estimated by Steiner \textit{et.al.}~\cite{bib29}.
}
\label{Fig3}
\end{figure}
% -----------------

Now, we select preferable RMF EoSs from the MR relation.
The MR relation has one-to-one correspondence to the EoS
through the Tolman-Oppenheimer-Volkov (TOV) equation~\cite{bib40}
\begin{align*}
 &\frac{dP}{dr}=-G\frac{M\epsilon}{r^2}
  \left(1+\frac{P}{\epsilon}\right)
  \left(1+\frac{4\pi Pr^3}{M}\right)
  \left(1-\frac{2GM}{r}\right)^{-1},
\\
 &\frac{dM}{dr}=4\pi r^2\epsilon,
\end{align*}
where $G$ is a gravitational constant
and $\epsilon$ is an energy density.
The NS has a crust region at low densities.
As an EoS of the crust region,
we use that of Miyatsu \textit{et.al.}~\cite{bib41}.

In solving the TOV equation,
the electron and the muon should be taken into account
to satisfy the charge neutral condition.
We treat the electron as a massless free Fermion and
the muon as a massive free Fermion.
If the number densities, $n_{\textrm{e}}$ and $n_{\mu^{-}}$, of
electron and muon are known,
the charge neutral condition is given by
\begin{eqnarray}
n_{\textrm{p}}=n_{\textrm{e}}+n_{\mu^{-}}
\end{eqnarray}
for the proton number density $n_{\rm p}$.
In the inner of NS, the $\beta$-equilibrium condition
is also satisfied:
\begin{align}
 \mu_{i}=b_{i}\mu_{\rm B}-q_{i}\mu_{\rm e}
\end{align}
for $i=$ p, n, e, $\mu^{-}$,
where $b_{i}$ ($q_{i}$) is the baryon number
(the electric charge) of particle $i$ and
$\mu_{\textrm{e}}$ is the electron chemical potential.
Solving the TOV equation numerically with the EoS that satisfies
Eqs. (9) and (10), we can get the MR relation.

Figure. 3 illustrates the MR relation calculated by MTEC EoS,
TM1 EoS and NL3 EoS.
The data on MR relation in Fig. 3
are taken from Refs~\cite{bib27, bib28, bib29}.
The maximum mass $M_{\textrm{max}}$ and radius $R_{\rm max}$
is tabulated in TABLE \ref{tb:table4}.
From Fig. 3, one can see that MTEC EoS is most consistent with
all the data, particularly in the $2M_{\rm sun}$ region.
TM1 EoS predicts a bit larger maximum radius,
but it considerably well reproduce the data of MR relation.
In NL3 EoS, the resulting $M_{\rm max}$ and $R_{\rm max}$
are inconsistent with the data of MR relation.
We therefore take MTEC and TM1 EoSs as the hadron-phase EoS
and construct the TPMa1--TPMa3, TPMb1--TPMb3.

%%%%% TABLE 4 %%%%%
\begin{table}[t]
\begin{center}
\caption{The maximum mass ($M_{\rm max}$) and radius ($R_{\rm max}$)
predicted from three RMF models.
The mass is normalized by the mass of sun $M_{\rm sun}$.
}
\begin{tabular}{c|c|c|c}
\hline \hline
 & MTEC & TM1 & NL3 \\ \hline
$M_{\textrm{max}}/M_{\rm sun}$ & 2.02 & 2.18 & 2.77 \\
$R_{\textrm{max}}$ (km) & 10.8 & 12.3 & 13.2 \\ \hline \hline
\end{tabular}
\label{tb:table4}
\end{center}
\end{table}
%%%%%%%%%%%%%%%%%%%

%%%%%%%%%%%%%%%%%%%%%%%%%%%%%%%%%%%%%%%%%%%%%%%%%%%%%%%%
\subsection{TRANSITON LINE OF TPMa1 AND TPMb1}
%%%%%%%%%%%%%%%%%%%%%%%%%%%%%%%%%%%%%%%%%%%%%%%%%%%%%%%%
We first consider the possibility of the hadron-quark phase transition
in the core of NS by using TPMa1 and TPMb1.
If the quark phase appears in the core of NS,
Eqs. (9) and (10) should be also imposed on the quark-phase EoS:
\begin{align*}
 & \frac{2}{3}n_{\rm u}-\frac{1}{3}n_{\rm d}=n_{\rm e}+n_{\mu^{-}},
\\
 & \mu_{\rm u}=\frac{1}{3}\mu_{\rm B}-\frac{2}{3}\mu_{\rm e},
\\
 & \mu_{\rm d}=\frac{1}{3}\mu_{\rm B}+\frac{1}{3}\mu_{\rm e},
\end{align*}
where $n_{\rm u}$ ($n_{\rm d}$) is the u-quark (d-quark) number density.
Which phase is realized is determined from the Gibbs criterion.

% ----- Fig.4 -----
\begin{figure}[t]%[H]
\begin{center}
\includegraphics[width=0.42\textwidth]{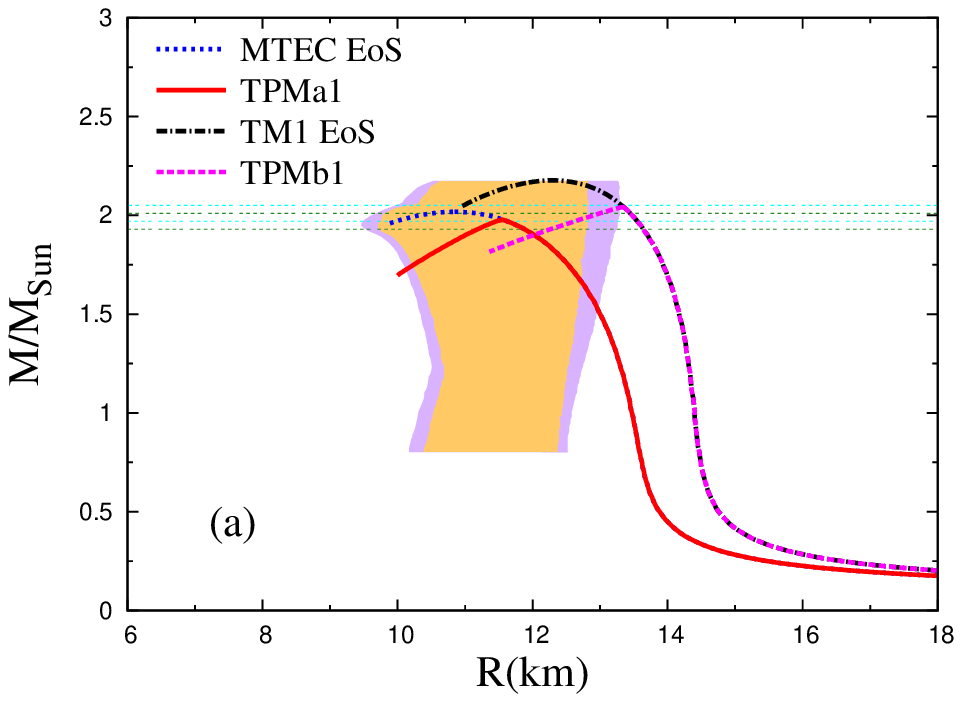}
\includegraphics[width=0.42\textwidth]{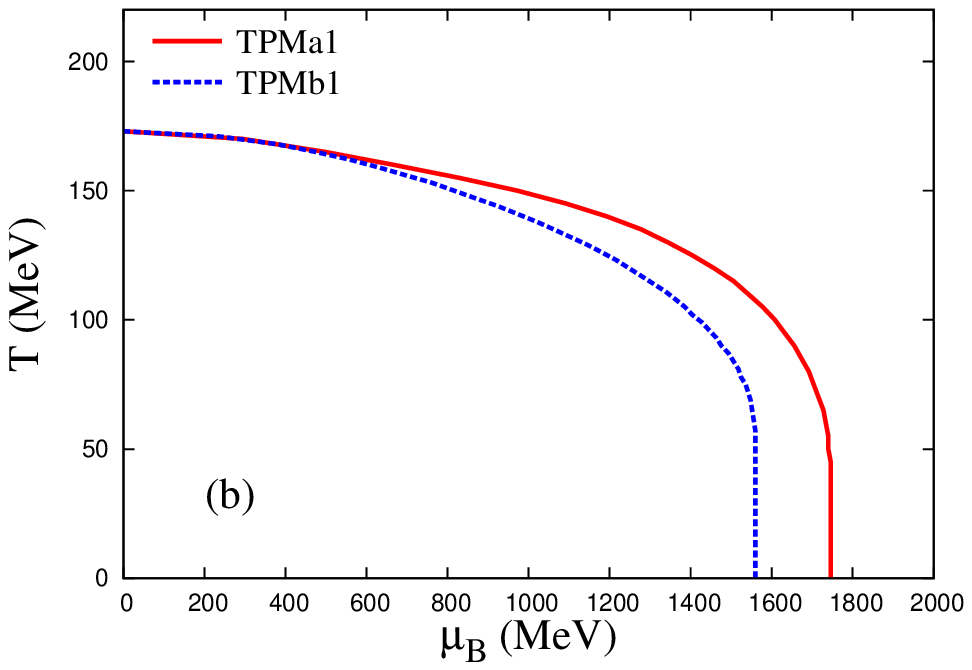}
\end{center}
\caption{The panel (a) shows the
MR relation calculated from the TPMa1 (solid), TPMb1 (dashed),
For comparison, the MR relation calculated from MTEC and TM1 EoSs
are also plotted.
In the panel (b), the hadron-quark phase transition lines
for TPMa1 and TPMb1 are drawn.
}
\label{Fig4a}
\end{figure}
% -----------------

% ----- Fig.5 -----
\begin{figure}[t]%[H]
\begin{center}
\includegraphics[width=0.42\textwidth]{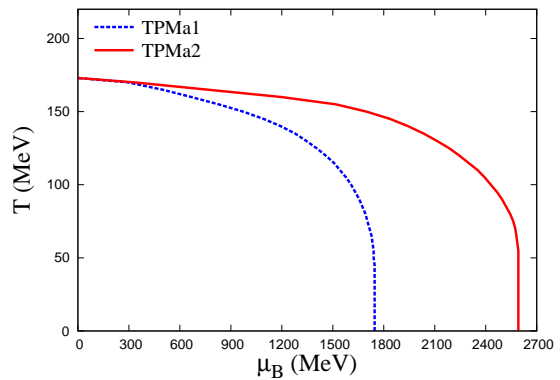}
\end{center}
\caption{
The hadron-quark phase transition line for TPMa1 and TPMa2.
}
\label{Fig5}
\end{figure}
% -----------------

In Fig. 4, the panel (a) shows the MR relations calculated with TPMa1 and TPMb1.
For comparison, the results calculated from MTEC and TM1 EoSs
are plotted.
In TPMa1, the quark phase appears
at $M=1.97M_{\rm sun}$ before reaching $M_{\rm max}=2.02M_{\rm sun}$
and consistent with the data on MR relation.
Also in TPMb1,
the quark phase emerges
at $M=2.04M_{\rm sun}$ before reaching $M_{\rm max}=2.17M_{\rm sun}$.

The panel (b) of Fig. 4 shows the hadron-quark phase transition line
in the $T$-$\mu_{\rm B}$ plane for TPMa1 and TPMb1.
The critical baryon chemical potential $\mu^{\rm c}_{\rm B}$ of the transition
at $T=0$ is 1750 MeV for TPMa1 and 1560 MeV for TPMb1.
If the $G_{\rm v4}(0)$ is positive,
the quark-matter EoS becomes stiffer and thereby
the predicted values of NS mass and $\mu^{\rm c}_{\rm B}$
are increasing.
Therefore, TPMa1 and TPMb1 yield the lower bound of
$\mu^{\rm c}_{\rm B}$ for each class of TPM
for the quark phase to take place in the core of NS.

% ----- Fig.6 -----
\begin{figure*}[t]%[H]
\begin{center}
\includegraphics[width=0.45\textwidth]{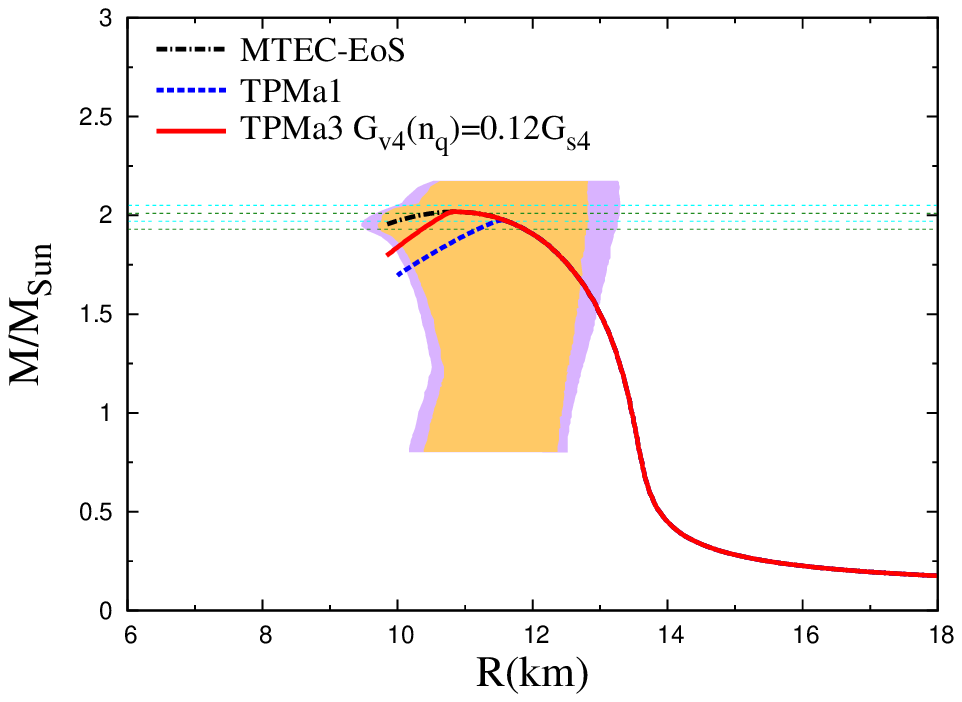}
\includegraphics[width=0.45\textwidth]{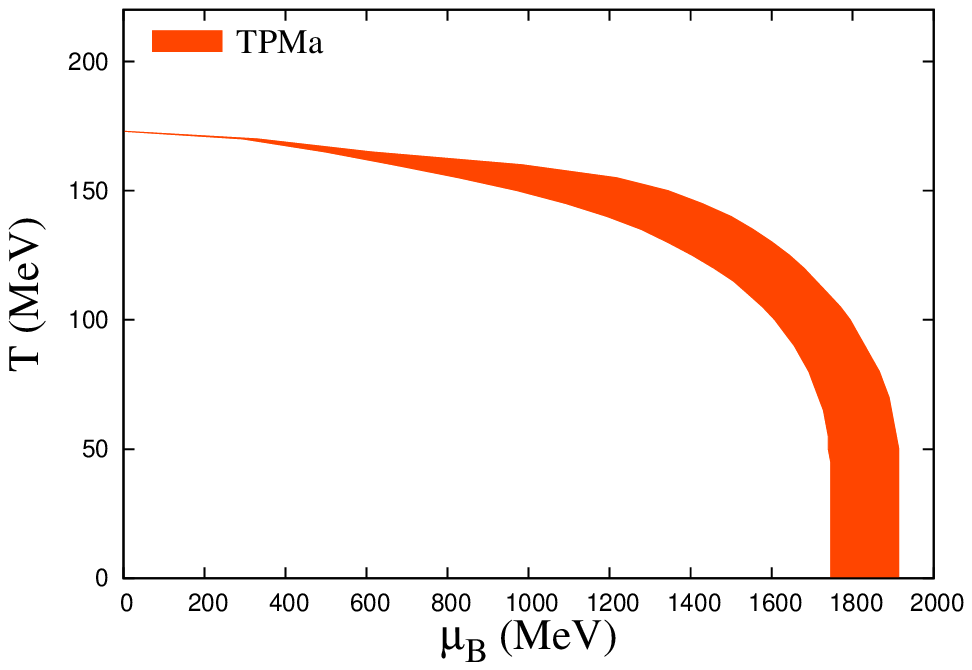}
\end{center}
\caption{
(Left panel) The MR relation calculated from TPMa3.
The results from TPMa1 and MTEC EoS are also plotted.
\\
(Right panel) The band of the hadron-quark phase transition line.
The upper (lower) bound of the band is calculated from
TPMa3 (TPMa1).
This region allows the quark phase appear in the core of NS.
}
\label{Fig6}
\end{figure*}
% -----------------

% ----- Fig.7 -----
\begin{figure*}[t]%[H]
\begin{center}
\includegraphics[width=0.45\textwidth]{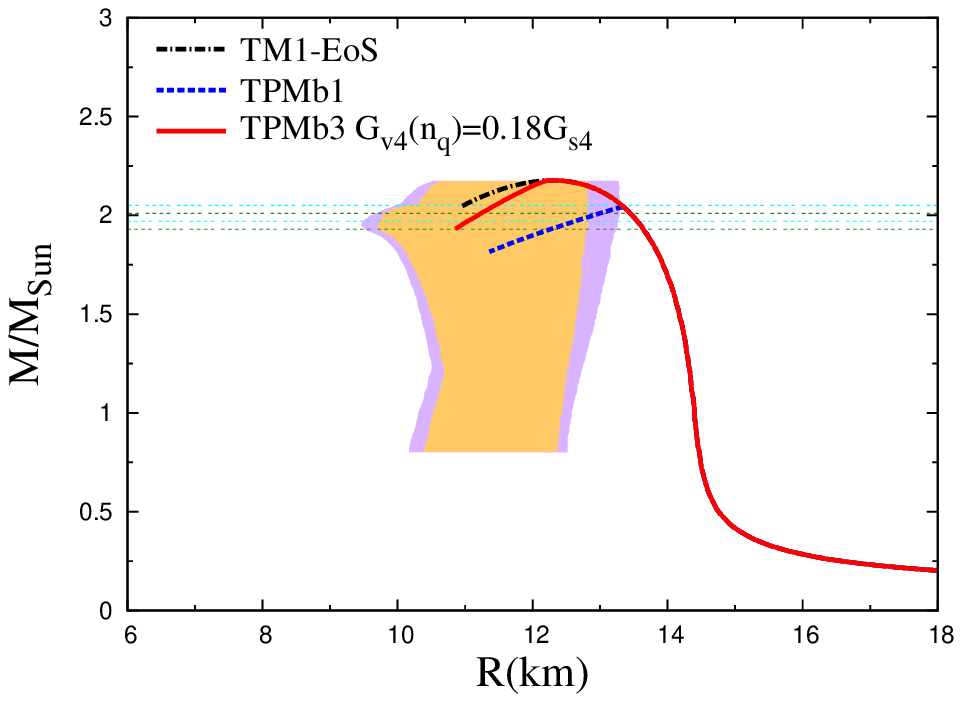}
\includegraphics[width=0.45\textwidth]{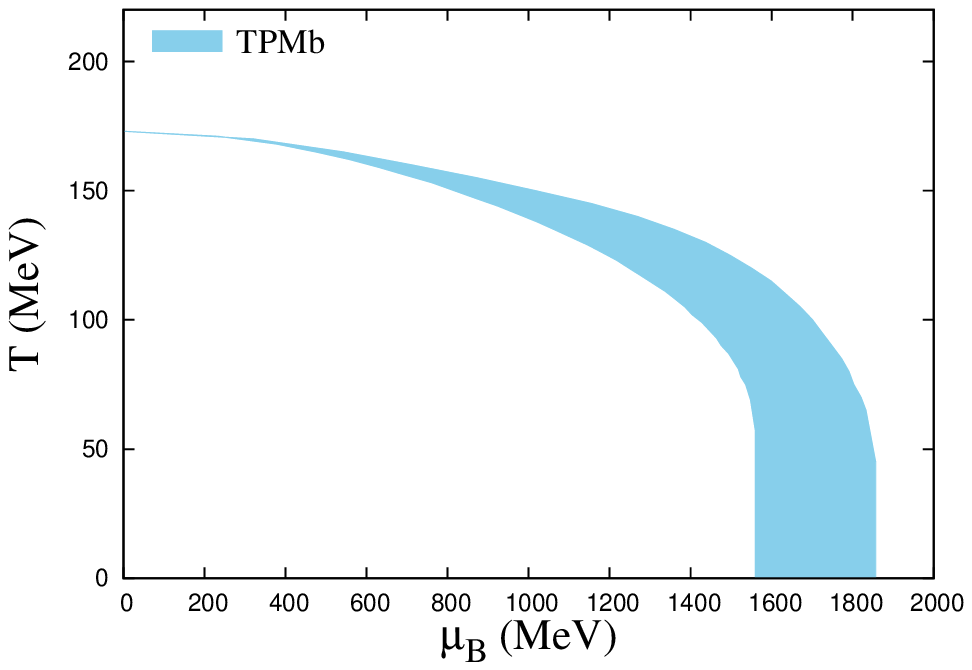}
\end{center}
\caption{
(Left panel) The MR relation calculated from TPMb3.
The results from TPMb1 and TM1 EoS are also plotted.
\\
(Right panel) The band of the hadron-quark phase transition line.
The upper (lower) bound of the band is calculated from
TPMb3 (TPMb1).
This region allows the quark phase appear in the core of NS.
}
\label{Fig7}
\end{figure*}
% -----------------

%%%%%%%%%%%%%%%%%%%%%%%%%%%%%%%%%%%%%%%%%%%%%%%%%%%%%%%%
\subsection{TRANSITON LINE OF TPMa2 AND TPMb2}
%%%%%%%%%%%%%%%%%%%%%%%%%%%%%%%%%%%%%%%%%%%%%%%%%%%%%%%%
Next, we consider TPMa2 and TPMb2
with $G_{\rm v4}(0)=0.36G_{\rm s4}$.
Figure. 5 illustrates the hadron-quark phase transition line
for TPMa1 and TPMa2.
One can see that the existence of $G_{\rm v4}(0)$
delays the transition toward higher $\mu_{\rm B}$.
The value of $\mu^{\rm c}_{\rm B}$ for TPMa2 is
2600 MeV and the corresponding density is 13$\rho_{0}$.
Such a density does not realize in the core of NS and
hence the quark phase does not appear in the core of NS for
TPMa2.

As for TPMb2, we find that
the hadron-quark phase transition line does not reach the $\mu_{\rm B}$ axis.
The reason is that the self interaction $(\omega^{\mu}\omega_{\mu})^2$
of $\omega$ meson more stabilizes
the hadron phase with respect to increasing $\mu_{\rm B}$,
while the vector-type four-quark interaction suppresses
the appearance of quark phase.
In fact, the quark phase is confirmed 
to never appear in the core of NS for TPMb2.

%%%%%%%%%%%%%%%%%%%%%%%%%%%%%%%%%%%%%%%%%%%%%%%%%%%%%%%%
\subsection{DENSITY DEPENDENCE OF $G_{\rm v4}$ AND
TRANSITON LINE OF TPMa3 AND TPMb3
}
%%%%%%%%%%%%%%%%%%%%%%%%%%%%%%%%%%%%%%%%%%%%%%%%%%%%%%%%
Finally, we consider TPMa3 and TPMb3.
In TPMa3 and TPMb3, the quark phase is described by
the EPNJL of type (3), that is,
the strength of vector-type four-quark interaction
depends on the quark number density $n_{\rm q}$
(See Eq. (\ref{eq:density-dependence})).
For TPMa3 (TPMb3), $\rho_{0}=0.153\ (0.145)\ \textrm{fm}^{-3}$ is used.
The form Eq. (\ref{eq:density-dependence}) ensures that
the interaction is invariant under the charge conjugation and
$G_{\rm v4}(n_{\rm q})$ is positive for any $n_{\rm q}$.
When $G_{\rm v4}(n_{\rm q})$ is negative, there is possibility that
vector meson masses calculated with the random-phase-approximation
becomes negative. Consequently, the $G_{\rm v4}(n_{\rm q})$ varies
in a range $0\le G_{\rm v4}(n_{\rm q}) \le G_{\rm v4}(0)=0.36G_{\rm s4}$.
We discuss the lower bound of $b$ by assuming that
the quark phase takes place in the core of NS.

The left panel of Fig. 6 shows the MR relation calculated with TPMa3.
In TPMa3, the quark phase appears at $M_{\rm max}=2.02M_{\rm sun}$
and $n_{\rm q}=7.2\rho_{0}$,
when the value of $G_{\rm v4}(n_{\rm q})$ is equal to $0.12G_{\rm s4}$.
This means that $0.12G_{\rm s4}$ is 
the maximum value of $G_{\rm v4}(n_{\rm q})$
for the quark phase to appear in the core of NS.
The corresponding value of $b$ is 0.001.
The right panel of Fig. 6 illustrates the hadron-quark phase transition line.
The lower bound of line is determined by the TPMa1 and
upper bound is the TPMa3 with $b=0.001$.
The values of $\mu^{\rm c}_{\rm B}$ lies in the
$1750 \textrm{MeV} \le \mu^{\rm c}_{\rm B}\le 1910$ MeV.
If the value of $\mu^{\rm c}_{\rm B}$ exists in this region,
the hadron-quark phase transition occurs in the core of NS. 
Note that the maximum value $\mu^{\rm c}_{\rm B}=1910$MeV is much smaller than 
$\mu^{\rm c}_{\rm B}=2600$MeV in TPMa2 shown in Fig.~\ref{Fig5}.

The left panel of Fig. 7 shows the MR relation calculated with TPMb3.
As for TPMb3, the quark phase appears at $M_{\rm max}=2.17M_{\rm sun}$
and $n_{\rm q}=6\rho_{0}$,
when the value of $G_{\rm v4}(n_{\rm q})$
is equal to $0.18G_{\rm s4}$, which is the maximum value of $G_{\rm v4}(n_{\rm q})$
for the quark phase appear in the core of NS in TPMb3.
The corresponding value of $b$ is 0.001 and
common between TPMa3 and TPMb3.
The right panel of Fig. 7 illustrates the hadron-quark phase transition line.
The lower bound of line is determined by the TPMb1 and
upper bound is the TPMb3 with $b=0.001$.
The values of $\mu^{\rm c}_{\rm B}$ lays in the 
$1560 \textrm{MeV} \le \mu^{\rm c}_{\rm B}\le 1860$ MeV.
The lower bound of $\mu_{\rm B}^{\rm c}$ is not same between
TPMa1 and TPMb1, but upper value for TPMa3 and TPMb3 is nearly equal.

%%%%%%%%%%%%%%%%%%%%%%%%%%%%%%%%%%%%%%%%%%%%%%%%%%%%%%%%%%%%%%%%%%%%%%%%%%%
%%%%% Summary
%%%%%%%%%%%%%%%%%%%%%%%%%%%%%%%%%%%%%%%%%%%%%%%%%%%%%%%%%%%%%%%%%%%%%%%%%%%
\section{SUMMARY}
In this paper, we constructed the TPM in which
the EPNJL model is used in the quark phase and
the RMF model is in the hadron phase.
To make the TPM reasonable,
we took LQCD data and NS observations as reliable constraints.
For the quark-phase model,
we determined the density-independent strength $G_{\rm v4}(0)$ of
vector-type four-quark interaction
from LQCD data on $n_{\rm q}/n_{\rm SB}$
in the $\mu_{\rm q}/T=0$ limit
with small error bars.
The obtained value is $G_{\rm v4}(0)=0.36G_{\rm s4}$ that
is a bit larger than our previous work.
For the hadron phase, we take
three RMF models; NL3, TM1 and MTEC.
We compared calculated MR relations with observed ones.
We found that MTEC is most consistent with the data
and TM1 is the second best, while NL3 is inconsistent.

We then take MTEC and TM1 for the hadron part of TPM and
considered six types of TPMs (TPMa1--a3 and TPMb1--b3)
that are combinations of the two types of hadron-phase EoS
and EPNJL of type (1)--(3).
For TPMa3 and TPMb3,
we introduced
the density-dependent strength $G_{\rm v4}(n_{\rm q})$ of
vector-type four-quark interaction and
assumed that the density-dependence is
described as a Gaussian form having the single parameter $b$.

The MR relation and hadron-quark phase transition line are
calculated for six TPMs.
As a result,
the hadron-quark phase transition occurs in the core of NS
when $1750 \textrm{MeV} \le \mu^{\rm c}_{\rm B} \le 1910$ MeV for
TPMa
and $1560 \textrm{MeV} \le \mu^{\rm c}_{\rm B} \le 1850$ MeV for
TPMb.
For both TPMa and TPMb, the corresponding minimum value of $b$ is
$b=0.001$.

\noindent
\begin{acknowledgements}
We thank G. M. Mathews, T. Kajino, J. Takahashi, and M. Ishii for useful discussions. 
J. S., H. K., and M. Y. are supported
by Grant-in-Aid for Scientific Research (No. 27-7804, No. 26400279, and No. 26400278)
from the Japan Society for the Promotion of Science (JSPS). 
\end{acknowledgements}

%%%%%%%%%%%%%%%%%%%%%%%%%%%%%%%%%%%%%%%%%%%%%%%%%%%%%%%%%%%%%%%%%%%%%%%%%%%%%%%%%%%%% References 
%%%%%%%%%%%%%%%%%%%%%%%%%%%%%%%%%%%%%%%%%%%%%%%%%%%%%%%%%%%%%%%%%%%%%%%%%%%%%%%%

\end{document}